\documentstyle[]{l-school}
\input{epsf}
\begin{document}
\markboth{P. Haensel}{SOLID INTERIORS OF NEUTRON STARS}
%
%
\title{SOLID INTERIORS OF NEUTRON STARS \\ 
        AND GRAVITATIONAL RADIATION}
%
\author{P. Haensel}
\institute{ N. Copernicus Astronomical Center\\ 
             Polish Academy of Sciences\\
                 ul. Bartycka 18, 00-716 Warszawa\\
		  Poland}
\maketitle
\centerline{\sl to be published in 
      ``Astrophysical Sources of Gravitational Radiation''}  
\centerline{\sl Les Houches 1995, eds. J.-A. Marck 
and J.-P.Lasota}
\vskip 5mm 
\newcommand{\dens}{{\rm g~cm^{-3}}}
\newcommand{\Msun}{M_\odot}
\newcommand{\Mac}{M_\odot/{\rm y}}
\section{INTRODUCTION}
Neutron stars are the densest stable stellar objects in the
present day Universe. For the typical masses  $M\sim 1 -
2~M_\odot$, they are expected to have radii $R\sim 10~$km, so
that their mean density  $\bar\rho\sim 10^{15}~{\rm
g~cm^{-3}}$, significantly higher than the normal nuclear
density, $\rho_0=2.5\times 10^{14}\dens$, characteristic of
interiors  of heavy atomic nuclei.
Neutron stars are formed as very hot objects, as 
an outcome  of
gravitational collapse of massive stellar cores at the endpoint
of their thermonuclear evolution, or - under specific conditions
 - in the gravitational
collapse of mass accreting white dwarfs. Newly born neutron
stars are very hot, with $T_{\rm initial}\sim 10^{11}~$K (which 
corresponds to 
$k_{\rm B}T\sim 10~$MeV). They cool rapidly due to neutrino 
emission from
their interiors, so that after one year their internal
temperature falls below $10^{9}~$K. At such temperatures, matter
in the neutron star interior is strongly degenerate (due to its
huge density), so that the effects of temperature on the
structure of neutron star, older than one year, can  be
neglected. 
Neutron stars are observed as radio pulsars, X-ray pulsars, and
as X-ray bursters. Since they contain matter of the density
ranging from a few ${\rm g~cm^{-3}}$ in the atmosphere to
$\widetilde> 
10^{15}~{\rm g~cm^{-3}}$ at the center, they should be
considered as `cosmic laboratories' for superdense matter.
Observations of neutron stars yield constraints on the theory of
dense matter which was used for constructing neutron star models
(an introduction to physics and astrophysics of neutron stars
can be found in \cite{ST83})

Neutron stars are massive ($M\sim 1 - 2~M_\odot$), compact
($2GM/Rc^2\sim 0.3$), and rapidly rotating (period of rotation
$P\sim 1~$ms for millisecond pulsars) - and therefore they can
be considered as potentially promising 
sources of continuous gravitational
radiation. The shortest observed period of rotation (for radio
pulsar PSR 1937+21) $P=1.56~$ms, which corresponds, for
$R=10~$km, to the equatorial velocity $v_{\rm equator}\sim
0.1 c$. The expected typical moment of inertia of neutron star
 is $I\sim 10^{45}~{\rm g~cm^2}$. Crucial for the emission of
continuous gravitational radiation 
 is the time dependence (in the laboratory frame) of
the quadrupole moment tensor, $Q_{\rm ij}$, 
of the stellar mass distribution. Let us denote the amplitude of
the time dependent part of $Q_{\rm ij}$ by $I\epsilon_{\rm
eff}$, where $\epsilon_{\rm eff}$ describes the deviations from
the axial symmetry (with respect to the rotation axis), and is
ususally referred to as {\it effective tri-axiality} of neutron
star \cite{ST83}. The mean power of the gravitational
radiation, emitted by rotating neutron star with non-vanishing
$\epsilon_{\rm eff}$, is proportional to $\epsilon_{\rm eff}^2$,
and strongly increases with rotation frequency, 
$\dot E_{\rm GR}\propto I^2 \epsilon_{\rm eff}^2 P^{-6}$ 
 \cite{ST83}. 
 
The non-vanishing value of $\epsilon_{\rm eff}$ appears only
under specific conditions. It can be sustained due to a partially
solid nature of neutron star (a completely fluid, stably
rotating body is symmetric with respect to rotation axis). As
discussed in Sections 6 - 8, the existence of the solid
interiors of neutron stars can lead to a non-zero 
$\epsilon_{\rm eff}$ in two cases: first, due to the appearance
of genuine tri-axiality, resulting from the presence of
`mountains' on the neutron star crust, or due to the neutron
star precession. Two other possibilities, not related to the
solid nature of neutron stars, and therefore not discussed in
these lectures, should be mentioned. The first possibility
corresponds to strong magnetic
field, with symmetry axis different from the rotation axis.
However, such magnetic field should be extremely strong
($B\widetilde > 10^{16}~$G), and should have quite 
special spatial structure to maximize 
$\epsilon_{\rm eff}$ (see the lecture of 
 S. Bonazzola and E. Gourgoulhon 
in this volume). 
 The second possibility, valid also for a
completely fluid star, is connected with appearance of
non-axisymmetric, gravitational radiation reaction  driven  
instabilities,  \cite{FriedIps92} 
(see also the lecture of R. Wagoner in this volume), or with the shape
instabilities, discussed in this volume in the lecture 
of S. Bonazzola and E. Gourgoulhon.
 However, both  gravitational radiation reaction 
and shape instabilities are expected  
 for a  rotation frequency
$\Omega=2\pi/P$ close to the mass-shedding limit, $\Omega\simeq
\Omega_{\rm MS}$, while observed $\Omega$'s are significantly lower
than $\Omega_{\rm MS}$.

An additional attractive feature of a (partially) 
solid neutron star with 
non-zero $\epsilon_{\rm eff}$ is  connected with the remarquable
stability of pulsar rotation. This could allow for long
integration times (years) of periodic gravitational radiation signal
from a pulsar with non-zero $\epsilon_{\rm eff}$. 

The existence of a {\it solid crust} of neutron star is well
established on the theoretical ground.  
Theory  predicts the
solid character of dense, cool (degenerate)  matter  for the
density  below $\rho_{\rm h}\simeq 10^{14}~{\rm g~cm^{-3}}$. 
The existence of the solid crust  is confirmed in a
rather convincing way by the observations of glitches in the
pulsar timing \cite{ST83}.
On the contrary, the possibility of existence of 
the {\it solid cores} of neutron
stars, with density  $\sim 10^{15}~{\rm g~cm^{-3}}$, is based on
rather exotic models of superdense matter, and should be
considered as interesting  but very uncertain. 
\section{FORMATION AND COMPOSITION OF THE SOLID CRUST}
In the present section we will restrict ourselves to the outer
layer of the neutron star crust, with  density up to a few times 
 $10^{13}~\dens$. In this  density regime, the volume occupied by
the atomic nuclei constitutes a small fraction of the total
volume and atomic nuclei, present in the matter, can be treated
as spherical. The properties of the bottom layers of the crust,
with $\rho\simeq 10^{14}~\dens$, within which the assumption of
sphericity of nuclei may break down, will be  discussed in Section
 3.
\subsection{Formation of the crust: high temperature scenario}
The initial temperature of the interior  of 
neutron star formed in gravitational
collapse is $T_{\rm initial}\sim 10^{11}~$K, which corresponds to
the typical energy of thermal motion of matter constituents
$k_{\rm B}T\sim 10~$MeV. At such temperature matter is a multicomponent
plasma of atomic nuclei $(A,Z)$, alpha particles $\alpha$,
neutrons $n$, protons $p$, and electrons and positrons $e^-$,
$e^+$. The baryon density of matter (number of baryons - in our
cases nucleons - in a unit volume) is given by $n_{\rm b}=n_{n}+n_{p}
 + 4 n_\alpha+n_A A$, where $n_{\rm i}$ is the number density of a
species `i', and where we assumed, for simplicity,
 that only one heavy nuclid  $(A,Z)$ is
present in the matter 
(this is a rather good approximation, see \cite{Burr84}). 
Charge neutrality of matter requires that $n_p
+n_{e^+} + 2 n_\alpha + Z n_A = n_{e^-}$. Under the conditions
prevailing in the interior of a newly born neutron star, matter
can be assumed to be in complete nuclear equilibrium (the rate
of all reactions is much higher than the rate of evolution of
neutron star interior). At given $n_{\rm b}$ and $T$, the
composition of matter
 is obtained from the
minimization of the free energy density $f=e-Ts$, 
where $e$ is the internal energy density, and $s$ is entropy 
per unit volume.
 Neutron star
cools rapidly, due to neutrino losses, and after one year
internal temperature is less than $10^{9}~$K. A standard
assumption is, that during the cooling process matter was always
close to the state of nuclear equilibrium, with composition
always given by the minimum of $f$. However, at $T<10^{9}~$K
thermal contribution to $f$ is negligible (for
$\rho>10^{6}~\dens$), and the $T=0$ approximation (ground state
of matter) is excellent. The composition of matter can be then
calculated from the requirement $e=minimum$, at fixed $n_{\rm b}$,
and under the condition of charge neutrality. Such a procedure
of determining the {\it ground state} of dense matter enables
one to calculate the composition and equation of state of the
outer envelope of neutron star, formed from the initial, very
hot state. The ground state of matter turns out to be a
body-centered-cubic  (bcc) lattice of nuclei $(A,Z)$, immersed in the
uniform background of electrons. The neutron fraction in nuclei, 
$N/Z$, 
increases with density, and above {\it neutron drip} density,
$\rho_{\rm ND}=4-6\times 10^{11}~\dens$, some of neutrons 
form a gas outside nuclei, so that for $\rho>\rho_{\rm ND}$
ground state of matter is a bcc crystal of nuclei, embedded in a
neutron gas, and permeated by a uniform background of electron
gas. 

\subsection{Formation of the crust: low temperature scenario}
Many neutron stars are (or were in the past) members of close
binary systems, in which the  second member is (was) 
an evolved, large  
radius (and therefore low surface gravity) star.  
Due to  its huge gravitational field, neutron star accretes
matter from its companion. Accretion of matter increases the
mass of neutron star, leads to formation of a new outer mantle, and
influences the temperature distribution within the neutron star
interior.  After the quasi-stationary thermal state of the
neutron star 
interior has been reached, the internal temperature profile can
be obtained from the balance of energy gains and losses. The
energy gains are due to heating by
infalling matter and to  work done by gravitational forces compressing
the stellar interior. The heat loss is due to photon emission  
from the neutron star 
photosphere and neutrino losses from the neutron star 
interior. The calculations, based on realistic neutron star
 models, lead
to conclusion, that for typical accretion rates in close binary
systems, $10^{-11}~\Mac<\dot M< 10^{-9}~\Mac$, internal
temperature is always lower than $10^9~$K
 \cite{Fuji84,Fuji87,Miralda90}. At such temperatures
the assumption of full thermodynamic equilibrium during the
formation of the new neutron star
 crust is not valid: the  energy of
thermal motion of nuclei, 
$E_{\rm therm}= {3\over 2}k_{\rm B}T$, is
so low compared to typical
Coulomb barriers, $E_{\rm Coul}$, that 
 nuclear reactions involving charged particles are practically 
 suppressed (blocked) (on the timescale of the
neutron star evolution). The pressure and density
in a given shell of the crust of accreting neutron star
increase, as the accretion proceeds. The local value of electron
chemical potential (Fermi energy) increases 
according to $\mu_e\propto (\rho Z/A)^{1/3}$,
 for
$\rho>10^7~\dens$. 
However, the composition of
the matter can change only via those reactions between matter
constituents, which are energetically possible under 
prevailing low
temperature conditions. The only reactions, which
can proceed under conditions prevailing in the neutron star crust
are thus: {\it electron capture}  on nuclei,
\begin{equation}
(A,Z)+e^-\longrightarrow (A,Z-1) +\nu_e\;,
\label{eq:eCap}
\end{equation}
{\it neutron emission} (if triggered by electron captures),
\begin{equation}
(A,Z) +e^- \longrightarrow (A-{k},Z-1) +\nu_e + {k}\cdot n\;,
\label{eq:nemiss}
\end{equation}
 as well as neutron absorption, and {\it pycnonuclear fusion} of
neighbouring nuclei, resulting from quantum tunnelling of the
Coulomb barrier due to quantum zero-point motion around the
crystal lattice sites,
\begin{equation}
 (A,Z) + (A,Z) \longrightarrow (2A-{k},Z) +{k}\cdot n\;.
\label{eq:pycno}
\end{equation}
The pycnonuclear regime of fusion 
(Greek {\it pycnos} - means dense) is to be contrasted with
thermonuclear one;  the
kinetic energy of nuclei  results there from the quantum effect of
localization, which at sufficiently high density  
dominates over the thermal one: $E_{\rm
zero~point}\gg E_{\rm thermal}$. The value of $E_{\rm
zero~point}$ increases with increasing density, which results in
a strong density dependence of the rate of pycnonuclear 
fusion \cite{ST83}. 

The crust of an accreting 
neutron star has a characteristic, time
dependent layer structure. Let us assume, that during time $t$ 
the accretion of matter onto the neutron star
 surface took place at
constant rate $\dot M$. After time $t$, the outer layer of 
neutron star of
mass $\Delta M=M_{\rm accr}=\dot M t$ has thus been 
replaced by the
accreted matter. This new shell of mass $\Delta M$ compresses
the `old' crust, laying below it. Each layer of the `old
crust' undergoes  compression, characterized by the compression
factor, equal to the ratio of its actual density, $\rho$ (at
time $t$),  to its initial, original density, $\rho_{\rm i}$ 
(before accretion). Let us characterize the original (initial) position of
the crust layer by the mass of  the layers laying above it,
$\Delta M_{\rm i}$, and let us denote the corresponding 
compression  factor by $C(\Delta M_{\rm i})
=\rho/\rho_{\rm i}$.
 Notice, that the pressure at the bottom of the layer is
proportional to the mass laying above it. 
Consider the case of $M_{\rm accr}=5\times 10^{-4}~\Msun$, which
corresponds to $t=5\times 10^6$~y at accretion rate $\dot M =
10^{-10}~\Mac$. In what follows,  we will quote values obtained
for a specific model of neutron star, of mass $1.4~\Msun$, with
liquid core ($\rho>\rho_{\rm h}$) calculated for the BJI equation
of state \cite{HZ90b}. For a layer,  laying originally at the
neutron star 
surface (for this layer $\Delta M_{\rm i}=0$), 
one gets then $C=10^{13}$: the density of this layer after
accretion of $\Delta M_{\rm accr}$ is $\sim 10^{13}~\dens$. 
However, due to huge
pressure gradient in the surface layers, the value of $C$
decreases rapidly with increasing $\Delta M_{\rm i}$. We get
$C(\Delta M_{\rm i}=5\times 
10^{-4}~\Msun)\simeq 2 $ (i.e.,
the density of this layer increase from $10^{13}~\dens$ 
 to $2\times 10^{13}~\dens$). 
Finally, we get $C(\Delta
M_{\rm i} = 10^{-3}~\Msun)\simeq 1$. The compression of the
deeper layers of  neutron star crust can be neglected, and so their
composition coincides with the original one.  

Let us summarize our results. If neutron star accreted $M_{\rm
accr}$, then the composition of its outer layer of mass $M_{\rm
accr}$  may be
expected to deviate widely from the original (ground state) one: this
layer of neutron star crust was formed under conditions
differing strongly from the `high temperature' scenario, described
in the preceding section. Its actual composition can be obtained
by following the evolution of matter, which was originally a
hydrogen rich plasma lost by the neutron star companion. Such a
program was fulfilled in \cite{HZ90a,HZ90b}. The differences with
respect to the ground state decrease gradually with increasing
density, and for 
 the crust layers laying below the outer envelope of the mass 
$2M_{\rm accr}$ 
 the approximation
of the ground state composition is very good. 
\subsection{Composition of the crust, melting temperature}
Composition of the crust is important for determining its
physical properties, such as melting temperature, elastic parameters,
and thermal and electrical conductivity. Within standard
approximation, the composition of neutron star crust will be
determined by giving the values of $A$
and $Z$  of the nuclid present in the matter and, above the
neutron drip point, also the mass fraction contained in  the 
neutron gas outside nuclei, $X_n$.  These parameters, for the
ground state of the crust, and for the case of accreted crust, 
are shown in Figs.~1,~2. As we see from
Fig.~1, the values of $Z$ and $A$ in accreted crust are
significantly smaller, than those characteristic of the ground
state of matter. The differences increase with increasing
density, and for $\rho> 10^{11}~\dens$, 
 $Z$ and $A$ in the accreted crust are 2-3 times
smaller than the ground state one. As seen from 
Fig.~2, neutron drip in an
accreted crust occurs at a somewhat higher density, than in the
case of the ground state of dense matter, but the values of
$X_n$  at higher densities are not dramatically different. 
 In both cases, matter undergoes {\it neutronization}: the proton
mass fraction decreases as the density
increases. 
\begin{figure}
\begin{center}
\leavevmode
\epsfxsize=9cm \epsfbox{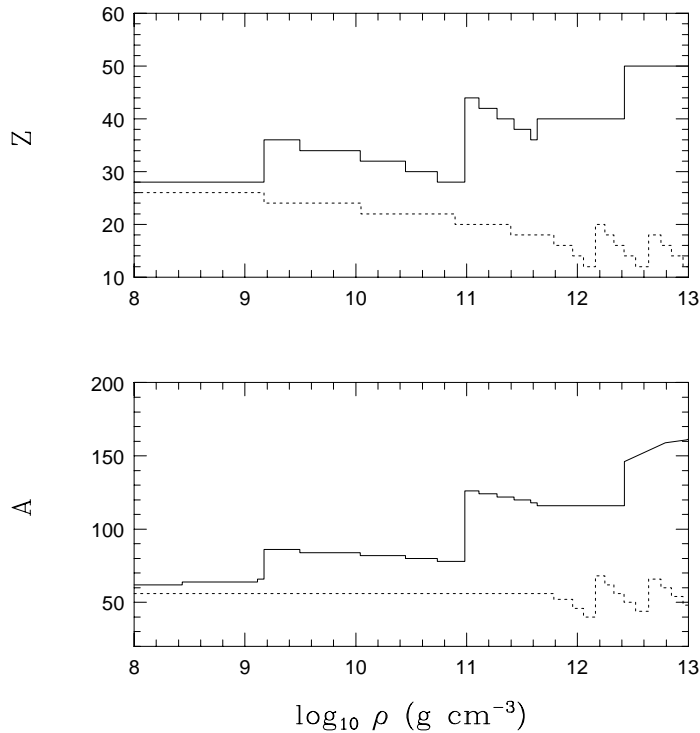}
\end{center}
\label{Fig1}
\caption[]{
The values of $Z$ (upper panel) and $A$ (lower panel) for the nuclids
present in dense matter for the ground state (solid line: based on
 [5] below neutron drip point 
and [4] above
neutron drip) and for the accreted crust 
(dashed line: based on
 [6]). 
}
\end{figure}
\begin{figure}
\begin{center}
\leavevmode
\epsfxsize=9cm \epsfbox{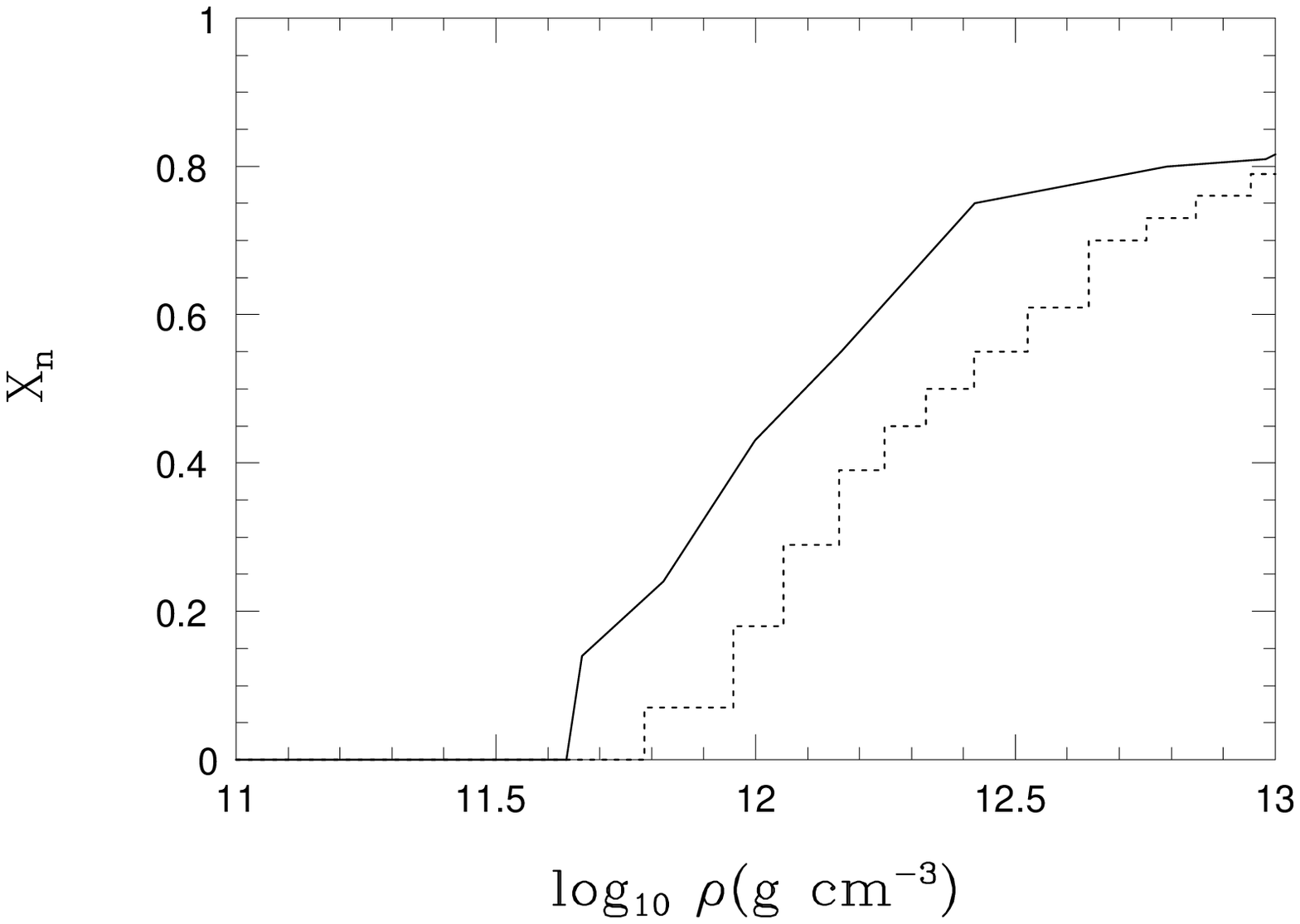}
\end{center}
\label{Fig2}
\caption[]{
Mass fraction contained in the neutron gas outside nuclei,
$X_n$, for the ground state of dense matter (solid line) and
accreted crust (dotted line)
}
\end{figure}

One notices characteristic features of the density dependence of
$A$ and $Z$,  represented in Fig.~1. They can be
understood in terms of the general properties 
of nuclei. Only  even-even
nuclei (even numbers of neutrons and protons) are present  - which
results from the action of the pairing forces between nucleons.
 In the
case of the ground state of dense matter, strong shell effects 
are clearly seen: this results from the fact, that closed-shells
nuclei are particularly strongly bound.  For density  below
$10^9~\dens$,  
increasingly neutron  rich  Ni isotopes are present,
 with $Z=28$ (closed proton shell). Then, up to the neutron drip
point, 
nuclids with closed neutron shells $N=50$ and $N=82$ are present
in the ground state of matter. Finally, above neutron drip,
closed $Z=40$ proton subshell and $Z=50$ proton 
shell are persistent, while
neutron shell effects are no longer visible (which is due to the
presence of the outer, unbound neutron gas). Let us notice, that
up to the density $\rho\simeq 10^{11}~\dens$ the composition of
the ground state of matter is determined by the experimental
data \cite{HPichon94} ! In the case of accreted matter, one
starts with $^{56}{\rm Fe}$ shell (produced from the electron
captures on 
$^{56}{\rm Ni}$, obtained at the considered
accretion rate  by the explosive burning of helium formed by
steady burning of accreted, 
originally hydrogen rich plasma). Then one has to follow 
 the evolution of this
shell during low temperature compression up to $\sim
10^{13}~\dens$ \cite{HZ90b}. 
\begin{figure}
\begin{center}
\leavevmode
\epsfxsize=7cm \epsfbox{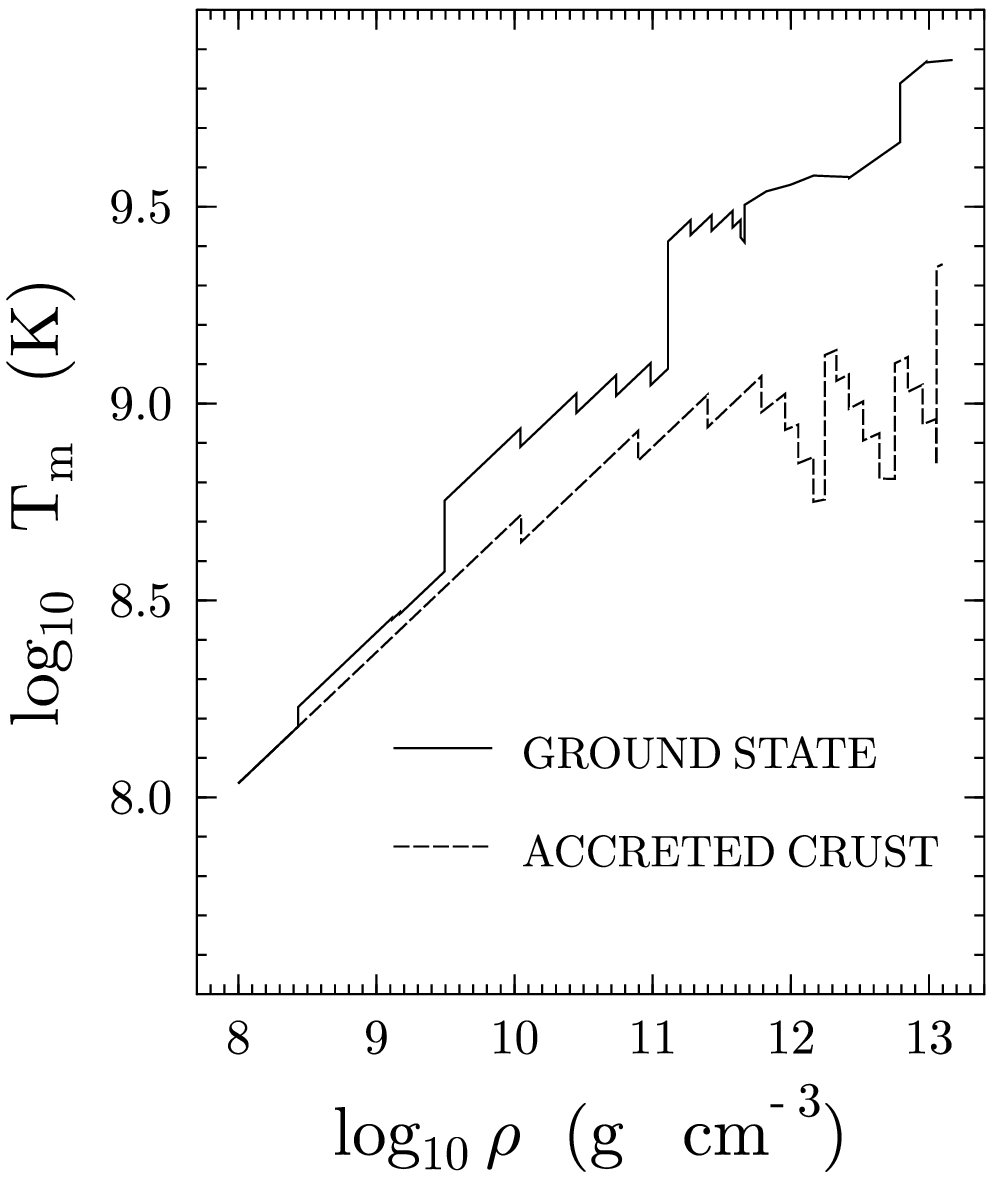}
\end{center}
\label{Fig3}
\caption[]{Melting temperature versus density, for the ground
state composition  of dense matter (solid line) and for the
accreted crust (dashed line)
}
\end{figure}

Up to now, we neglected the effect of finite temperature of
neutron star crust. While at $T<10^9~$K the effect of $T$ on the
composition of matter is negligible, the value of $T$ decides 
whether, at a given density,   
dense plasma is in the liquid or solid (crystal) phase.
Crystallization of the plasma is due to Coulomb forces between its
constituents, which lead to localization of ions (nuclei). The
importance of Coulomb energy, as compared to thermal
energy of nuclei, is measured by the Coulomb
coupling parameter,
\begin{equation}
\Gamma={Ze^2\over a k_{\rm B}T}\;,
\label{eq:Gamma}
\end{equation}
where $a$ is the radius of the sphere of the volume `per one
nucleus', 
 $1/n_A$ 
(volume of the Wigner-Seitz  (W-S) cell, $V_{\rm cell}$
\footnote{
The Wigner-Seitz cell is a useful object in the plasma physics;
in our approximation it is a sphere containing one nucleus and
$Z$ electrons, to be  electrically neutral. To a very good
approximation, the energy of a plasma is a sum of energies of
Wigner-Seitz cells.}
). 
 The many-body
calculations of the thermodynamic properties of dense
 one-component plasma show that, at fixed composition, solid phase
of the plasma (bcc crystal) is energetically preferred over the
liquid one 
 for $\Gamma>\Gamma_{\rm m}\simeq 178$ \cite{Slatt82}. This
corresponds to the melting temperature
\begin{equation}
T_{\rm m}=1.32\times 10^6\; Z^2\; 
               \left(\rho_9\over A_{\rm cell}\right)^{1\over
3}~{\rm K}\;,
\label{eq:Tm}
\end{equation}
where $A_{\rm cell}$ is the number of nucleons in the
Wigner-Seitz cell, and $\rho_9=\rho/(10^9~\dens)$. For
$\rho<\rho_{\rm ND}$, we have  $A_{\rm cell}=A$, 
while for $\rho>\rho_{\rm
ND}$ the total number of nucleons in the W-S cell is 
 $A_{\rm cell}=A+n_n V_{\rm cell}$. 

The melting curve  $T=T_{\rm m}(\rho)$ separates the liquid and solid
regions in the density - temperature plane. The melting curves
for the ground state of dense matter, and for the accreted
crust, are shown in Fig.~3. The difference between the
two melting curves are significant for $\rho>10^9~\dens$, and
above neutron drip point $T_{\rm m}({ ground~state})\gg
T_{\rm m}({accreted~ crust})$. This reflects strong $Z$-
dependence of melting temperature, combined with rather weak
dependence on $A_{\rm cell}$ and $\rho$.
\section{THE BOTTOM LAYERS OF THE CRUST: EXOTIC 
         NUCLEAR SHAPES AND TOPOLOGIES}
For $\rho>\rho_{\rm ND}$ nuclei can be considered as droplets of
`nuclear matter', of nearly constant density $\rho_{\rm
drop}\simeq \rho_0=2.5\times 10^{14}~\dens$, immersed in a less
dense neutron gas. 
So, above neutron drip density, 
nucleons are present in two
coexisting phases: gaseous (neutron gas,  or  G-phase) 
and liquid (nuclear
matter, or  L-phase). As long as the fraction of the volume
occupied by nuclei is a small fraction of the total volume
of the system, droplets of nuclear matter can be treated as
spherical - such a  shape minimizes the surface energy of the
droplets. At $T=0$, the energy density at a given baryon density
$n_{\rm b}$  can be decomposed as
\begin{equation}
e(n_{\rm b},~{parameters})=e_{{\rm vol},A} + e_{\rm surf} +
e_{\rm Coul} + e_e + e_n\;,
\label{eq:drop}
\end{equation}
where $e_{{\rm vol},A}$ is the bulk (volume) part of the
contribution from nuclei (droplets), $e_{\rm surf}$ is the
contribution, resulting from the existence of the interface
between the neutron gas and nuclear matter (surface energy), and
$e_{\rm Coul}$ is the contribution  from Coulomb
interactions of charged constituents of matter. Finally, $e_e$
is the  energy density of electron gas, which permeates both
phases of nucleonic matter, and $e_n$ is the contribution of
neutron gas outside nuclei (droplets). The description 
of nuclei embedded in neutron gas
is usually done using {\it compressible liquid drop model} of
atomic nuclei. Within this model, the problem of determination
of the ground state of matter at some fixed $n_{\rm b}$ 
 is reduced to minimization of 
$e(n_{\rm b},~{parameters})$ with respect to all parameters,
under the condition of charge neutrality. Let us notice, that
the parameters of `droplets' are modified with respect to the
zero pressure case (characteristic of terrestrial nuclei) due to
pressure exerted by neutron gas (which results in compression 
of nuclei); the presence of
neutron gas influences also the surface energy term in Eq.
(\ref{eq:drop}). 

The fraction of volume occupied by the denser L-phase increases
with matter density: however, even at $\rho=10^{13}~\dens$ we have
$V_{\rm L}/V=2\times 10^{-2}$ \cite{NV73}, and the approximation
of spherical nuclei can be expected to be  valid. The situation changes,
however, when we consider matter of the density close to
$\rho_0$, where $V_{\rm L}$ constitutes a sizable fraction of
$V$: then, the spherical droplets may no longer minimize the
energy density of the system. Such a possibility was considered
long time ago by Baym, Bethe and Pethick \cite{BBP70}, who
showed that at  
$V_{\rm L}/V=1/2$ nuclei turn `inside out', and
at higher $V_{\rm L}/V$ spherical bubbles of neutron gas in
nuclear matter are energetically preferred. 

The problem of actual nuclear shapes for $\rho_0/3 < \rho < \rho_0$ was
recently reconsidered by Lorentz et al. \cite{LRP93}. Using the
compressible liquid drop model, these authors optimized the
value of $e(n_{\rm b},~ {parameters},~{shape})$ with
respect to the {\it shape} of the interface between the 
L and G-phase. The shapes they considered were: in  
three dimensions (3-D) - spherical
droplets of L-phase in homogeneous G-phase, 
and spherical bubbles of G-phase in
homogeneous L-phase; 
in 2-D - cylindrical rods of L-phase in homogeneous G-phase,
and cylindrical tubes filled with G-phase in homogeneous L-phase. 
Finally,
considered 1-D shapes were alternating plates of G and L-phases.
The sequence of (first-order) phase transitions, obtained in
\cite{LRP93} for a specific compressible liquid drop model of
nuclei, 
starts at $\rho=1.06\times 10^{14}~\dens$, where $V_{\rm
L}=0.09$, $X_n=0.76$ and $X_p=0.03$, by a transition to `rods'.
Then, with increasing density, transitions from `rods' to
`plates', from `plates' to `tubes', and from `tubes' to
`bubbles' take place. Finally, at $\rho_{\rm h}=1.6\times
10^{14}~\dens$ matter becomes a homogeneous mixture of neutrons,
protons and electrons ($V_{\rm L}=V$, $X_p=0.03$, $X_n=0.97$).
The value of $\rho_{\rm h}$ corresponds to the bottom of the solid
crust as determined in  \cite{LRP93}. This value depends somewhat on the
assumed parameters of the compressible liquid drop model, used
for the description of the the L-phase, and on the $n-n$
interaction model, 
relevant for the description of the G-phase. 

\section{CRUST CONTRIBUTIONS TO STELLAR MASS  AND MOMENT OF INERTIA}
In order to calculate the contribution of neutron star crust to
neutron star  parameters, such as mass and moment of inertia,
let us consider the equation of hydrostatic equilibrium for a
non-rotating neutron star,
\begin{eqnarray}
{{\rm d}p\over {\rm d}r} &=& - {Gm\rho\over r^2}
 \left(1 +{4\pi r^3 p\over m c^2}\right)
\left(1 + {p\over \rho c^2}\right)\Lambda\;,
\nonumber \\
{{\rm d} m\over {\rm d}r} &=& 4\pi r^2 \rho\;,
~~~~~~~~~~~~~~~~\Lambda=\left(1-{2Gm\over r c^2}\right)^{-1}\;.
\label{eq:TOV}
\end{eqnarray}
For sufficiently massive neutron star ($M\widetilde> 1.2~\Msun$), the
crust is thin, and contains only a small fraction of the mass of
neutron star. For a shell at radius $r$ within the crust, 
both  $(R-r)/R$ and $(M-m)/M$ are typically few times $10^{-2}$.
Moreover, pressure within the crust is small compared to energy density,
$p/\rho c^2 < 10^{-2}$. All these conditions enable us to
rewrite Eq.(\ref{eq:TOV}) within the crust in the approximate form,
\begin{equation}
{{\rm d}p\over {\rm d}r}=- {GM\rho\Lambda(r)\over r^2}\;,
~~~~~~~~~~~~~~{{\rm d} m\over {\rm d}r} = 4\pi r^2 \rho\;,
\label{eq:TOVa}
\end{equation}
which can be rewritten, with our approximation,  as
\begin{equation}
{{\rm d}p\over {\rm d}m}=- {GM\rho\Lambda(R)\over 4\pi R^2}\;.
\label{eq:TOVcr}
\end{equation}
Integration of the above equation implies, that the mass 
of the layer of the crust is proportional to the pressure at its
bottom,
\begin{equation}
\Delta M (p_{\rm bot})\simeq 
{4 \pi R^4\over G M \Lambda (R)}\; p_{\rm bot}\;.
\label{eq:Mbot}
\end{equation}
Similar approximations, applied to the moment of inertia for
slow, rigid rotation, lead to the formula \cite{LRP93}
\begin{equation}
\Delta I(p_{\rm bot})\simeq 
{2\over 3}\Delta M(p_{\rm bot}) R^2
\left( 1- {2GI\over R^3 c^2}\right)\Lambda(R)\;,
\label{eq:Ibot}
\end{equation}
where $I$ is the total moment of inertia of the star. 

Combining these equations with equation of state for the crust,
one can see that  the innermost layer of the crust, containing 
`exotic nuclei', discussed in the preceding section, yields 
as much as half of the total crust  contributions to $M$ and $I$ ! 
For a $1.4~\Msun$ neutron star, 
$I_{\rm crust}$ and $M_{\rm crust}$ constitute  $\sim 1\;\%$ 
of  $I$ and $M$, respectively  (this agrees 
well with exact results reported in \cite{LRP93}). Both $M_{\rm
crust}/M$ and $I_{\rm crust}/I$ decrease with increasing $M$.

The value of $\rho_{\rm h}$, quoted in this lecture, deserves an
additional comment. Before the appearance of the Lorentz et al.
paper \cite{LRP93}, the commonly accepted value of $\rho_{\rm
h}$, based on an old calculation of
Baym et al. \cite{BBP70}, 
was some 50\% higher. Consequently, the present results for
$I_{\rm crust}$ and $M_{\rm crust}$ are about two times smaller,
than those obtained using the `old' value of $\rho_{\rm h}$ 
\cite{LRP93}.

\section{SOLID CORES}
The appearance of crystal structure for $\rho<\rho_{\rm h}$ is
due to electromagnetic (Coulomb) interactions. Possible
existence of {\it solid cores } in the liquid interior of
neutron stars would be due to specific features of the {\it
strong interactions} between nucleons. At supranuclear density,
localization of nucleons, characteristic of solid neutron star
cores, discussed in \cite{PS76,TT76,TT77,TT78,KW90,KW95}, 
 results in the increase of the mean kinetic energy,
$\Delta e_{\rm kin}$. In order to make the crystal structure
energetically preferred over the liquid one, the gain in the
potential energy, $\Delta e_{\rm pot}$, should more than
compensate the increase of kinetic energy: $\Delta
e_{\rm pot}<-\Delta e_{\rm kin}$. Calculations were based on
variational approach: for an assumed nuclear hamiltonian, the
structure of the ground state is usually  determined from $e(n_{\rm b},
{parameters})=minimum$. Transition from liquid to solid takes
place, if minimization on the crystalline (periodic) structure yields lower
$e$ than that obtained for homogeneous liquid.

The possibility of solidification of pure neutron matter at
sufficiently high density, due to
strong repulsive short-range neutron-neutron interaction, was
pointed out in the early 70-ties. However, subsequent calculations, based
on more reliable solutions of the nuclear many-body problem, did
not confirm these suggestions (see \cite{BP75} for the history
of the neutron solid). In the mid-seventies, 
 several authors suggested  existence of solid structures due to
appearance of the neutral-pion-condensate (\cite{PS76},
\cite{TT76}, \cite{TT78}). 
 In all cases, solid structures were
preferred because of the strong tensor component in the
nucleon-nucleon interaction, which correlated nucleon spins with
spatial directions. The specific crystalline structure depended
on the assumed nucleon hamiltonian, while the 
spatial periodicities were
simply related to the wave-length of the $\pi^0$ condensate,
$\lambda_{\pi^0}$. In the calculation, reported
in \cite{PS76}, the crystalline structure was a 3-D cubic
lattice of nucleons, with  lattice spacing
 $a={1\over 2}\lambda_{\pi^0}$. In
the paper of Takatsuka and Tamagaki \cite{TT78}, dealing with
pure neutron matter  with $\pi^0$ condensate, the energy was
minimized by a `1-D solid', consisting  of alternating layers of
neutrons, with opposite spin orientations, the distance between
the layers being  given by 
$d={1\over 2}\lambda_{\pi^0}$. It is more appropriate to
call such a structure a 1-D liquid crystal (detailed review of
the work of the Kyoto group on the crystal-like structures in
pion-condensed neutron star matter can be found in \cite{Japs93}). 
In all cases,
crystallization took place at $\rho=3-5\rho_0$. 

A distinctly diferent model of solid neutron star core was proposed by
Kutschera and W{\'o}jcik \cite{KW90,KW95}. 
As these authors show, for
sufficiently low $X_p<0.05$, protons may be treated as `impurities' in
neutron matter, and their behavior is determined by their
interaction with neutrons, $V_{np}$. At the densities
considered, $V_{np}$  becomes more repulsive with increasing
density, and this results in the decrease of neutron density
around a $p$-impurity. For some nucleon-nucleon interactions,
the crystalline structure appears at $\rho>3-4~\rho_0$, 
with protons {\it localized} in the potential wells 
around the neutron density minima. 
%

%
\section{DEFORMATION, ELASTIC STRAIN AND MOUNTAINS}
In contrast to the liquid phase of dense matter, the solid phase
can support a non-vanishing shear and  can be a site of
elastic strain.  We begin the present section 
with a brief summary of the
  basic concepts of the theory of elastic media. In
what follows, for the sake of simplicity we will restrict
ourselves to the standard, Newtonian  version of 
the theory of elasticity
 \cite{LLelast}. A consistent 
general-relativistic theory of elastic media can be formulated;
this was done by Carter and Quintana \cite{CQ72}.

The state of thermodynamic equilibrium (ground state at $T=0$,
if thermal effects are neglected) corresponds to specific
distribution of (equilibrium) positions of ions within the
neutron crust, ${\bf 
r}=(x_1,\; x_2\;, x_3)$. In what follows, we will neglect thermal
effects. Such a state corresponds to the minimum of the energy 
density,  $e=e_0=minimum$, and to
vanishing elastic strain. The {\it deformation} of an element of
the crust with respect to the ground state configuration of ions
implies a displacement of ions into new positions 
${\bf r'}={\bf r}+{\bf u}$, where 
${\bf u}={\bf u}({\bf r})$ is the 
{\it displacement} vector. This is accompanied by the
appearance of {\it elastic strain} 
(i.e., forces, which tend to return the system to the
equilibrium state), and {\it deformation energy density} 
$\Delta e=e-e_0$. Uniform translation, corresponding to 
${\bf r}$-independent displacement field ${\bf u}$ 
does not contribute
to $\Delta e$, and the  real (genuine) deformation is described by the
(symmetric) {\it strain tensor} 
\begin{equation}
{
u_{ik}={1\over 2}\left(
{\partial u_i\over \partial x_k} + 
{\partial u_k \over \partial x_i} 
\right)\;.
}
\label{eq:uik}
\end{equation}
The non-zero elastic strain contributes to the stress tensor of
matter, which for zero strain has the form $-{\breve p}\delta_{ij}$, and
which in the presence of elastic strain is given by
\begin{equation}
\sigma_{ij}=-{\breve p}\delta_{ij} + \sigma_{ij}^{\rm strain}\;.
\label{eq:sigmaij}
\end{equation}
In the case of an isotropic solid, which should be a reasonable
approximation to the {\it macroscopically averaged} neutron star
crust (consisting presumably of small, disordered on macroscopic scale,
 anisotropic domains), the elastic properties are determined by
two elasticity parameters: the compression modulus $K$, and the
shear modulus, $\mu$. 
 After the deformation, the volume of
an element of matter changes according to 
${\rm d}V'=(1+u_{jj}){\rm d}V$, 
where summation is understood over the repeated indices.  
The diagonal strain tensor,
$u_{ik}=a\delta_{ik}$,  
corresponds to pure compression.
 Compression term modifies the equation of state of the crust,
adding an elastic contribution to the pressure ${\breve p}$,
Eq.(\ref{eq:sigmaij}). 
The elastic  {\it shear } contribution to the stress
tensor reads
\begin{equation}
\sigma_{ik}^{\rm shear}=
 \mu\left( 2 u_{ik} -{2\over 3} u_{jj}\delta_{ik}\right)\;.
\label{eq:sigmaSTR}
\end{equation}

In the case of $u_{jj}=0$ we deal with a
pure {\it shear} deformation of matter element, with
$\sigma_{ik}^{\rm shear} = 2 \mu u_{ik}$. 
In general, elastic
strain involves both compression and shear.

For a specific model of solid neutron star crust, the value of
$\mu$ can be expressed as a function of matter density and
composition. For a bcc lattice structure below $\rho_{\rm ND}$
one gets \cite{BP71}
\begin{equation}
\mu=0.295 Z^2 e^2 n_A^{4\over 3}
=0.024 \left({Z\over 26}\right)^{2\over 3} p_e\;,
\label{eq:muNUM}
\end{equation}
where $p_e$ is the electron pressure. For $\rho<\rho_{\rm ND}$
we have $p\simeq  p_e$, so that $\mu\ll p$. In view of the fact,
that considered deformations will be small, the elastic effects
can be always treated as small perturbation of a fluid (i.e. zero
elastic strain) configuration. For 
$\rho>\rho_{\rm ND}$, one
has to take into account the presence of a neutron gas outside
nuclei, but the inequality 
$|\sigma_{ij}^{\rm strain}|\ll p$ is still
valid. The shear modulus of the neutron star crust, for the
ground state composition of dense matter and for the accreted
 crust, is plotted versus density in Fig.~4.

\begin{figure}
\begin{center}
\leavevmode
\epsfxsize=9cm \epsfbox{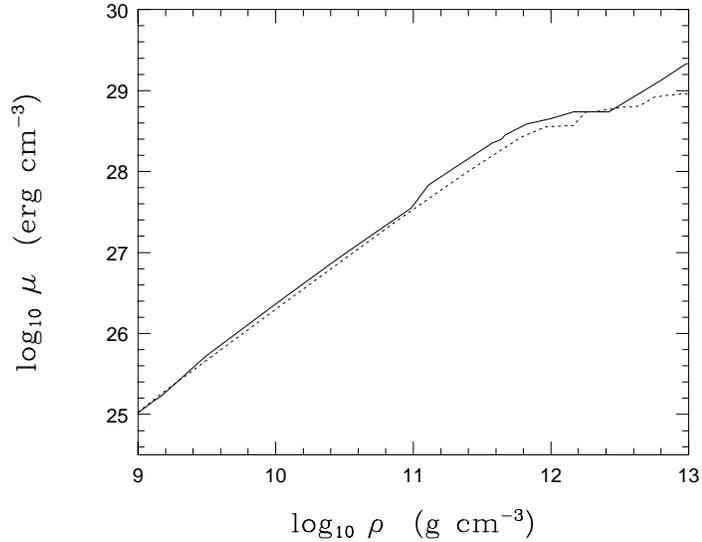}
\end{center}
\label{Fig4}
\caption{Shear modulus of neutron star crust versus density.
Solid line - ground state of cold dense matter. Dotted line -
accreted neutron star crust.}
\end{figure}

 In the case of exotic nuclear shapes, predicted for the density
interval  ${1\over 3}\rho_0< \rho < \rho_{\rm h}$, 
numerical  estimates of $\mu$ are not available, and in practice
one is thus forced to use 
the extrapolation of the lower density (bcc coulomb
lattice) results.

In the case of solid pion-condensed core, shear modulus has been
estimated in \cite{PPS76}, \cite{TT88}. In both cases one gets
an estimate  $\mu_{\rm core}\sim 10^{-2}-10^{-1}~p$, so that
 the elastic stresses of a hypothetical pion-condensed solid 
core  could contribute significantly to the total stress 
tensor of dense
matter. Actually, the perfect solid structure considered in
\cite{TT88} had a one-dimensional character (alternating
layers); however, the macroscopically averaged core should be
expected to behave as an isotropic solid. In the case of bcc
lattice of localized protons, considered in \cite{KW95}, we
expect that $\mu_{\rm core}<<p$, because of the small proton
fraction in dense matter (the smallness of proton fraction was
actually necessary for proton localization). 

The presence of elastic strain modifies the equation of
mechanical equilibrium of neutron star. The equation of
mechanical equilibrium  of an element of neutron star reads,
for Newtonian gravitation,
\begin{equation}
{\partial \sigma_{ij}^{\rm shear}\over \partial x_j} 
+ \rho g_i - {\partial p\over \partial x_i} = 0\;,
\label{eq:hydeq}
\end{equation}
where $g_i$ is the local value of gravitational acceleration.
The presence of $\sigma_{ij}^{\rm shear}$ allows for the
permanent deformations of the neutron star crust. These
deformations, supported by elastic stress, can lead to {\it
non-sphericity} of non-rotating neutron star, and {\it
non-axiality} of rotating neutron star. Such deformations -
expected to be small (see below) - bear some resemblance to
`mountains' on surfaces of solid planets. The upper bound on the
height of the  `mountains' on the neutron star crust will be
discussed in the final part of this section.

Another interesting possibility, stemming from the partially
solid structure of neutron star,  is that of {\it precession}.
Precession is possible in the case of the non-equal principal
moments of inertia, e.g., 
when $I_{xx}=I_{yy}\neq I_{zz}$.
Precession of partially solid neutron stars, which produces 
a non-vanishing value
of effective tri-axiality parameter, $\epsilon_{\rm eff}$, will
be discussed in the subsequent section.

Up to now, we restricted ourselves to Newtonian theory of
elasticity. General relativistic theory of elasticity has been
formulated in \cite{CQ72}. In its version suitable for
application to isotropic solids, relativistic theory involves
two material parameters $K$ and $\mu$ - which in the limiting
case  of weak gravity coincide with their Newtonian
counterparts. The theory formulated in \cite{CQ72} has been
applied to fully relativistic description of partially
solid,  rotating neutron stars in \cite{CQ75}. 

`Mountains' on the neutron star crust could be formed as a
result of the neutron star crust-quakes. Such crust-quakes  are
expected to be due to cracks (disruption of the crystal structure)
 in the
crust, implied by sufficiently large stresses (exceeding the
critical values that the crust can support), and arising, e.g., 
during the slowing down of pulsar rotation. Let us consider a
partially solid neutron star, rotating uniformly at the angular
frequency $\Omega=2\pi/P$ around the $z$
- axis, which is assumed to be a principal axis of the moment of
inertia tensor. Let us 
characterize the deformation of the neutron star
by the asymmetry in the principal moments of inertia, 
$(I_{xx}-I_{yy})/I=\epsilon$, where $I=I_{zz}$. Such a 
rotating, non-axisymmetric neutron star, will be a source of
pulsating gravitational radiation, of angular frequency
$\Omega_{\rm GR}=2\Omega$. The mean power of gravitational waves
can be calculated using the formula derived in Section 16.6 of \cite{ST83},
\begin{equation}
\dot E_{\rm GR} = 
{32\over 5} {G\over c^5} I^2 \epsilon^2 \Omega^6\;.
\label{eq:GRmount}
\end{equation}
This formula can be used in order to confront the hypothesis of
non-vanishing $\epsilon$ with pulsar timing. Emission of
gravitational radiation implies  additional loss of the kinetic energy
of the pulsar, and the slowing down of its rotation, according
to $\left(\dot E_{\rm kin}\right)_{\rm GR}=-\dot E_{\rm GR}$. In
view of the fact, that $\dot E_{\rm kin}=I\Omega\dot\Omega$,
this implies corresponding contribution to $\dot P$, 
which will be denoted by 
$(\dot P)_{\rm GR}$. Clearly,
the slowing down, implied by gravitational radiation should be
less that the observed one, $\left(\dot P\right)_{\rm GR}< \dot
P_{\rm observed}$. This implies an observational upper bound on
the value of $\epsilon$,
\begin{equation}
\epsilon < \epsilon_{\rm max}=3\times 10^{-9} 
\left({P\over 1~{\rm ms}}\right)^{3\over 2} 
\left({\dot P\over 10^{-19}}\right)^{1\over 2}\;,
\label{eq:epsobs}
\end{equation}
where we assumed  a `canonical' value of $I=10^{45}~{\rm
g~cm^2}$. The values of $\epsilon_{\rm max}$ for the Crab pulsar
 and for two selected millisecond pulsars, together with pulsar timing
parameters, are given in Table {\ref{Table1}}. 
\begin{table}
\caption{Timing parameters and $\epsilon_{\rm max}$ for three
selected radio pulsars}
\begin{tabular}{cccc}
&&\\
\hline\hline
&&\\
  Pulsar:  & 0437-47 & 1957+20 & Crab (0531+21)\\
&&\\
\hline
&& \\
&& \\
$P$(ms) & 5.76& 1.61 & 33.4\\
&&\\
$\dot P$ ($10^{-19}~{\rm s~ s~{-1}}$)& 1.2 & 0.17 & $4.2\times 10^6$\\
&& \\
d ({\rm kpc}) & 0.14 & 1.53 & 2.5 \\
&&\\
$\epsilon_{\rm max}$ & $3\times 10^{-8}$ & $2\times 10^{-9}$ & 
$1.2\times 10^{-3}$ \\
&& \\
\hline\hline
\end{tabular}
\label{Table1}
\end{table}

 The choice of pulsars in Table~{\ref{Table1}} deserves a comment.
The Crab pulsar was chosen as a good example of a young,
seismologically active pulsar (glitches observed in the Crab
pulsar timing are thought to result from neutron star
crust-quakes).  The remaining  two pulsars
are the millisecond ones; short rotation periods are preferred for an
efficient radiation of gravitational waves, 
Eq. (\ref{eq:GRmount}). 
 The PSR 0437-47 is  very close, 
 while PSR 1957+20 has the lowest value of
$\dot P$, and the second shortest period  of 
all observed radio pulsars: its timing yields the most stringent
upper bound on $\epsilon$. 
\section{SHAPE AND ENERGY OF ROTATING, PARTIALLY SOLID NEUTRON STAR}
The existence of solid regions in the interior of neutron star
(well established solid crust, speculative   solid core) opens
new possibilities for the structure and dynamics of rotating
neutron stars. Partially solid neutron star can be axially
non-symmetric, and can precess. In what follows, we will
describe these properties of partially solid neutron stars with
a simple `one parameter' model of Baym and Pines \cite{BP71}
(hereafter referred to as BP).
This model is  Newtonian, and therefore enables one a very simple
discussion of kinematics and energetics of rotating neutron 
star. Of
course, neutron stars are relativistic stellar objects, with
$2GM/Rc^2\sim 0.3$, and  the model of Baym and
Pines is an approximation of reality. A consistent, general
relativistic approach, based on the exact, relativistic theory
of elasticity, was formulated by Carter and Quintana
\cite{CQ75}(hereafter referred to as CQ). Connection between BP
and CQ parameters will be 
discussed at the end of this section.
\subsection{Liquid neutron star}
Consider rigid rotation of a  {\it liquid} neutron star around
the $z$-axis, with angular velocity $\Omega$. 
Its stationary configuration will be axially
symmetric with respect to $z$-axis which will,  at the same
time, be one of the three principal axes of the moment of inertia.
Due to rotation, $I_{xx}=I_{yy}<I_{zz}$. The deviation of
the shape of the star with respect to the spherical one is
described  by the `oblateness parameter' 
\begin{equation}
\epsilon = {I_{zz}-\bar{I}\over \bar{I}}\;,
\label{eq:epsIdef}
\end{equation}
where $\bar{I}=(I_{xx}+ I_{yy}+I_{zz})/3$. The total energy of the star is
an even function of $\epsilon$, and for $\epsilon \ll 1$, and at
 fixed total angular momentum $L$ is given by
\begin{equation}
E=E_*+{L^2\over 2I_{zz}} + A\epsilon^2\;,
\label{eq:Eeps}
\end{equation}
where $E_*$ is the total energy of a non-rotating configuration,
and where terms higher than quadratic have been neglected.
Stationary configuration corresponds to $E=minimum$, at
$L=I_{zz}\Omega=const$. This yields the expression for the
value of oblateness parameter of a liquid star,
\begin{equation}
\epsilon_\Omega=
{\Omega^2\over 4A}\left({\partial I_{zz}\over \partial \epsilon}
\right)_{\Omega=0}\simeq {\Omega^2 I_*\over 4 A}\;,
\label{eq:epsOm}
\end{equation}
where $I_*$ is the moment of inertia of a non-rotating star.
For the most rapid pulsars ($P=1.6~$ms) the typical value of
$\epsilon_\Omega$ is a few times $10^{-2}$ (assuming the
 mass $M\simeq 1.4~\Msun$) \cite{PPS76}; the exact value of
$\epsilon_\Omega$ depends both on the equation of state of dense
matter, and on the neutron star mass. 
\subsection{Rotation and elastic strain}
In general case, the solid regions of rotating neutron star are
a site of {\it elastic strain}, and in particular, they are
characterized by  non-vanishing shear stresses. The elastic
shear contributes to the local energy density of the solid
region of neutron star, with $e_{\rm shear}={1\over
2}\sigma_{ik} u_{ik}=\mu u_{ik}u_{ik}$. The contribution of
shear stress to the total energy of the star is in the BP model
represented by a simple expression,
\begin{equation}
E^{\rm shear}=B(\breve\epsilon - \epsilon)^2\;.
\label{eq:Eshear}
\end{equation}
The above formula represents a very important ingredient of the
BP model. The oblateness $\breve\epsilon$ corresponds to `reference
relaxed configuration' with zero shear (actually, with zero
elastic strain), consisting of the same number of baryons $N$.
This configuration will be referred to as $\breve{\cal C}$. The
most natural  assumption is that $\breve{\cal C}$ was the
configuration of neutron star just after the solidification of
the crust - so that its structure coincided, 
to a very good approximation,  with that of a fluid
star. The `elastically relaxed' configuration was rotating at a
 well defined $\breve\Omega$. By definition, 
$E^{\rm shear}=0$ for $\Omega=\breve\Omega$. 

The one-parameter formula of the BP model, which includes the
effect of elastic shear on the rotating configuration, reads
\begin{equation}
E=E_* + {L^2\over 2I} + A\epsilon^2 + 
B (\epsilon - \breve\epsilon)^2\;,
\label{eq:EepsShear}
\end{equation}

where by definition of configuration $\breve{\cal C}$,
\begin{equation}
\breve\epsilon 
={\breve\Omega^2\over 4 A}\;
\left({\partial I_{zz}\over \partial \epsilon}\right)_{\Omega=0}\;.
\label{eq:epsrelax}
\end{equation}

Stationary, rotating configuration with a fixed $L$ corresponds to
$E=minimum$. This yields the equilibrium value of $\epsilon$,
\begin{equation}
\epsilon=
{\Omega^2\over 4(A+B)} 
\left({\partial I_{zz}\over \partial\epsilon}\right)_{\Omega=0}
+{B\over A+B}\;\breve\epsilon\;.
\label{eq:epsAB}
\end{equation}
All effects  of elasticity enter through the parameter $B$,
another important parameter is the oblateness of the `relaxed
reference configuration, $\breve\epsilon$. Simple estimates of
$B$ for neutron star models with $M\simeq 1.4~\Msun$ with liquid
interior yield $B\sim 10^{-5}\;A$ \cite{PPS76}: the effect of
elasticity of the solid crust on the dynamics of rotating
neutron star can thus 
be treated as a small perturbation. In view of
$B\ll A$ it is suitable to introduce a small dimensionless 
`rigidity' parameter,
which measures the relative importance of elastic shear:
$b=B/A$. According to \cite{PPS76}, $b_{\rm crust}\sim 10^{-5}$.
 From Eq.(\ref{eq:epsAB}), the contribution of the elastic strain
to the oblateness of rotating neutron star is given by 
$(\epsilon)_{\rm elast}\simeq b_{\rm crust}\breve\epsilon$.
In the hypothetical case of a {\it solid core} in the
pion-condensed neutron star interior, the effect is expected to
be orders of magnitude larger, $b_{\rm core}\sim
10^{-3}-10^{-2}$ \cite{PPS76}. 

General relativistic version of 
one-parameter formula for partially solid,
rotating neutron star was derived by Carter and Quintana
\cite{CQ75AnnPhys}. Instead of making expansions in terms of 
the oblateness parameters
$\epsilon$, $\breve\epsilon$, Carter
and Quintana use $\Omega^2$ and $\breve\Omega^2$ as expansion
parameters. 
 In the slow rotation limit, 
the Newtonian parameters $A$ and $B$ 
are related to the relativistic parameters $P_*$ and
$Z_*$ of ref.\cite{CQ75AnnPhys}, by
\begin{equation}
P_*={I_*\over 2(A+B)}\;,
~~~~~~~Z_*={B I_*\over 2A (A+B)}\;,
\label{eq:PZ_AB}
\end{equation}
where $I_*$ is the moment of inertia of a spherical, 
non-rotating configuration.  
\section{OBLATENESS AND PRECESSION}
Even if originally fluid  neutron star rotated around its
symmetry axis, this situation might change after the star became 
partially solid. Let us consider rotating, partially solid
neutron star. Let the instantaneous axis of rotation be 
defined by  the
unit vector ${\bf n}_\Omega$. The deformation of neutron star
results from the combined action of centrifugal forces  and
elastic stresses within the neutron star crust. Both effects can
be treated as small perturbations. Let us switch-off deformation
due to rotation; the resulting {\it static} configuration will
still be deformed, due to elastic stresses within the neutron
star crust. The symmetry axis  of such a static but deformed
`reference' neutron star will be assumed 
to be given by the unit vector ${\bf
n}_0$. In the case of partially solid neutron star the vectors 
${\bf n}_\Omega$ and ${\bf n}_0$, 
may differ: 
this might result from the previous action of
a torque, which  tilted the  symmetry axis of the solid
component of the rotating body with respect to the rotation
axis. 
Such  torques, which could have acted on neutron star crust
 in the past, could be exerted by the neutron star magnetic
field,   by the tidal forces from a close companion, or by a
passing-by neutron star. Tilting could also result from a series
of `asymmetric' crust-quakes. 
In the general case, the unit vectors 
${\bf n}_\Omega$ and ${\bf n}_0$
will not be fixed in space, but will rotate around the fixed
direction determined by the angular momentum vector ${\bf L}$.
The rotation of ${\bf n}_\Omega$ around ${\bf L}$ corresponds to
well known {\it precession} of a solid body, typical 
 for the case
when the rotation axis does not coincide with any of the three
principal axes of the body moment of inertia. 
Keeping only terms linear in
oblateness parameters, we may write the inertia tensor of
rotating neutron star as \cite{PinSha72,AlpPin85}
\begin{equation}
I_{ij}=I_*\left[
(1-{1\over 2}\epsilon_\Omega - {1\over 2}b\breve\epsilon)\delta_{ij}
+ {3\over 2}\epsilon_\Omega n_\Omega^i n_\Omega^j 
+{3\over 2} b\breve\epsilon n_0^i n_0^j
\right]\;.
\label{eq:Iij}
\end{equation}
The angular momentum of the star is given by
$L_i=I_{ij}\Omega_j$. The three vectors ${\bf L}$, ${\bf
n}_\Omega$ and ${\bf n}_0$ are coplanar \cite{PinSha72}.

For a completely liquid star, we would have $b=0$, and we could
take  ${\bf n}_\Omega$ as  $z$-axis. This would give
\begin{equation}
I_{ij}^{\rm liquid}=I_*\left[
(1-{1\over 2}\epsilon_\Omega)\delta_{ij}
+ {3\over 2}\epsilon_\Omega\;\delta_{i3}\;\delta_{j3}
\right]\;,
\label{eq:IijLIQ}
\end{equation}
which would imply
\begin{eqnarray}
I_{xx}^{\rm liquid}=I_{yy}^{\rm liquid}&=&I_*(1-{1\over 2}\epsilon_\Omega)\;,
\nonumber\\
I_{zz}^{\rm liquid}&=&I_*(1+\epsilon_\Omega)\;.
\label{eq:IliqXYZ}
\end{eqnarray}
In the above equations, $I_*$ is the moment of inertia of a
spherical, non-rotating neutron star. 

In the linear approximation, the deformations described by
the oblateness parameters
 $\epsilon_\Omega$ and $b\breve\epsilon$ can be
treated independently. From the Euler equations, one finds that
the angular velocity vector (as calculated in the solid body
reference system) precesses around the direction
of ${\bf n}_0$, with precession frequency
\begin{equation}
\Omega_{\rm prec}={3\over 2} b \breve\epsilon\Omega\;.
\label{eq:Om.prec}
\end{equation}
The free precessional frequency of a partially solid neutron
star is thus determined by internal stellar structure and
distribution of internal elastic strain. 

While it might appear, that  the free precession 
of partially solid neutron stars is 
 a rather common phenomenon, observations of radio pulsars
give only one (and debatable) example 
of the possible presence of precession
in the pulsar timing. Analysing timing data
for the Crab pulsar, 
spanning the time interval of five years,  Lyne et al.
\cite{Lyne88} find  `quasi-sinusoidal' deviations from the
regular pulsar slow-down, with an approximate period of approximately
20 months. It should be stressed,  that  originally Lyne et al.
preferred  the
interpretation of it in 
terms of the so called `Tkachenko oscillations' of the
superfluid interior of neutron star. Further quantitative 
considerations indicated, however, the  relevance of the  
 solid interior of neutron star for explaining 
the specific period of modulations \cite{LynGrah90}.
 Precession of the solid crust remains thus one of possible
explanations \cite{LynGrah90}. 
Applying our precession model for explaining  observed
quasi-sinusoidal 
deviations, we would get $(b\breve\epsilon)_{\rm Crab}\sim
10^{-10}$. 
\section{GRAVITATIONAL RADIATION FROM PRECESSING PULSARS}
\subsection{Mean power of gravitational radiation}
Precession of a deformed pulsar implies radiation of
gravitational waves. The mean power of gravitational radiation
is calculated using the general formula,
\begin{equation}
\dot E_{\rm GR}={1\over 5}{G\over c^5} 
\left< 
{{\rm d}^3 {\cal I}_{jk}\over {\rm d}t^3}\;
{{\rm d}^3 {\cal I}_{jk}\over {\rm d}t^3}
\right>\;,
\label{eq:GRprec}
\end{equation}
where ${\cal I}_{ij}$ is traceless mass quadrupole moment in laboratory
(observer's) system \cite{ST83}. In the lowest order in $\epsilon$'s, terms
proportional to $\epsilon_\Omega$ do not contribute to the
relevant time derivatives, and therefore when calculating
${\cal I}_{ij}$ we may put $\epsilon_\Omega=0$. The calculation gives
\cite{ST83}:
\begin{equation}
\dot E_{\rm GR}={2\over 5}{G\over c^5} 
(I_1-I_3)^2_{\epsilon_\Omega=0}
\Omega^6 \theta^2\;,
\label{eq:GRprec2}
\end{equation}
where $\theta$ is the `wobble angle' (the angle between the
reference (`solid body') symmetry axis ${\bf n}_0$ 
and (fixed in laboratory, inertial reference frame) 
direction of ${\bf L}$; we
assumed that $\theta\ll 1$. In  
Eq. (\ref{eq:GRprec2}), $I_1$ and
$I_2$ are principal moments of inertia in the neutron star
(body) reference system ($I_1=I_{xx}$, $I_3=I_{zz}$).  The 
frequency of the gravitational waves is {\it approximately}
equal to the pulsar frequency, $\Omega$, namely 
$\Omega_{\rm GR}=\Omega + {3\over 2}b\breve\epsilon\Omega$ 
\cite{ZimSzed79,AlpPin85}. In our case, we have
\begin{equation}
I_1-I_3={3\over 2}b\breve\epsilon I_*\;,
\label{eq:I1-I3}
\end{equation}
so that the formula for $\dot E_{\rm GR}$ can be written in the
form
\begin{equation}
\dot E_{\rm GR}={32\over 5}{G\over c^5} 
I^2 \Omega^6 \epsilon_{\rm eff}^2\;,
\label{eq:GRprec3}
\end{equation}
where the {\it effective tri-axiality} is 
\begin{equation}
\epsilon_{\rm eff}={3\over 8} b\breve\epsilon \theta\;,
\label{eq:epsEFF}
\end{equation}
and $I\equiv I_*$. 
\subsection{Estimating $b$, $\breve\epsilon$, and $\theta$}
For a neutron star of mass $M\simeq 1.4~\Msun$ the existing
estimates give  $b_{\rm crust}\sim 10^{-5}$ \cite{PPS76}. 
In the hypothetical
case of a more massive neutron star with a solid pion condensed
core, a very uncertain estimate is  $b_{\rm core}\sim
10^{-3}-10^{-2}$ \cite{PPS76}. 

Let us first calculate the value of $\epsilon$ as a
function of angular frequency of radio pulsar. Using
Eq.(\ref{eq:epsOm}), we get
\begin{equation}
\epsilon\simeq\epsilon_\Omega = 
10^{-3}\;\left({I\over 10^{45}~{\rm g~cm^2}}\right)\;
\left({A\over 10^{53}~{\rm erg}}\right)\;
\left({P\over 10~{\rm ms}}\right)^{-2}\;.
\label{eq:epsOmNUM}
\end{equation}
Neutron star, born with some initial period of rotation, 
$P_{\rm init}$, slows down  
during its life as a radio pulsar. In
Table {\ref{Table2}} we give numerical estimates of initial
and present oblateness parameters, $\epsilon_{\rm init}$ and 
$\epsilon$ for the most rapid millisecond pulsar, PSR
1937+21, and for the Crab pulsar. 
\begin{table}
\caption{Initial and present oblateness parameters of two pulsars}
\begin{tabular}{ccc}
&&\\
\hline\hline
&&\\
  Pulsar:  & 1937+21 & Crab (0531+21)\\
&&\\
\hline
&& \\
&& \\
$P$ & 1.56 ms & 33 ms\\
&&\\
$P_{\rm init}$ & 1 s (10 ms) & 17 ms\\
&&\\
$\epsilon_{\rm init}$ & $10^{-7}$ ($10^{-4}$) & $3\times 10^{-4}$\\
&&\\
$\epsilon_\Omega$ & $4\times 10^{-2}$ & $9\times 10^{-5}$\\
&&\\
&&\\
\hline\hline
\end{tabular}
\label{Table2}
\end{table}
In both cases we assumed $I=10^{45}~{\rm g~cm^2}$ and
$A=10^{53}~{\rm erg}$. Initial period of the Crab pulsar was
calculated using the magnetic dipole formula for pulsar
slowdown \cite{ST83}. In the case of the PSR 1937+21, we
considered two cases. The millisecond pulsars 
 are believed to be previously 
radio-dead, old pulsars ($P\sim$~seconds) , 
revived  by accretion of matter in close binary
system. In the first case, we assumed that the old pulsar 
 (predecessor of PSR 1937+21) had in
the beginning of the accretion era small deformation, equal to
$\epsilon_\Omega$ at the beginning of this era, 
when $P=P_{\rm
init}=1~$s. 
We assumed additionally, that this initial configuration was
elastically relaxed (no elastic strain). Notice, that 
in the case of the Crab
pulsar the initial configuration was hot and liquid, while in 
 the case
of PSR 1937-21 the pulsar was assumed to be  
sufficiently old to be
elastically relaxed. 

In principle, in the case of PSR 1937+21, we could contemplate a
rather extremal situation, when the pulsar keeps the initial
oblateness from its `first birth'. Such a situation, which would
require a very high resilience of the stellar solid, would lead
to a Crab-like initial oblateness, which is indicated within the
brackets. Notice however, that during the accretion era the old
pulsar had to accrete some $\sim 0.1~\Msun$, in order to spin-up
to $P\sim 1~$ms. In view of this, the `old crust', with large
initial oblateness, was presumably `molten' by compression, and
transformed into the outer layer of the liquid interior. It
seems thus, that in order to keep the `first birth' oblateness,
the millisecond pulsar would have to contain a {\it solid core}.

The value of the `wobble angle', $\theta$, 
 is limited by the critical
strain, which the solid crust can support. The crust breaks down
(fractures) if the shear exceeds some critical value. In terms
of oblateness parameters, the condition for disruption of the solid is 
$|\epsilon - \breve\epsilon|>\Theta_{\rm crit}$, 
where
$\Theta_{\rm crit}$ 
is called `critical angle' (notice that
shear deformation changes angles within the  crystalline
structures) \cite{AlpPin85}. 
 A non-zero wobble  angle
implies an elastic strain within the solid region of the star. 
For typical terrestrial solids, 
$\Theta_{\rm crit}\sim 10^{-5}-10^{-4}$. 
This would imply 
$\breve\epsilon\simeq \epsilon_\Omega$ 
(a stringent condition in
the case of millisecond pulsars). Such a low $\Theta_{\rm crit}$
would limit the wobble angle to
$\theta\widetilde< 10^{-3}$ \cite{AlpPin85}. 
Let us notice, that  for a {\it perfect
Coulomb} lattice - which seems to be better suited
as a model of neutron star crust - calculations give 
$\Theta_{\rm crit}\sim 10^{-2}-10^{-1}$, 
which would imply
$\breve\epsilon\simeq \epsilon_{\rm init}$, 
and no limitations
on $\theta$ ! 
 (An interesting discussion of `elastic strength' of neutron star's 
crust,  
referring also to terrestrial experiments 
on `near Coulomb' metallic microcrystals, can be found in 
\cite{Ruderman92}.) 
 However, neutron star crust should expected to be
very far from the `perfect crystal' structure, 
and therefore we may expect
a significantly lower value of $\Theta_{\rm crit}$. In any case,
it is difficult to point out a mechanism, which could build up a
wobble angle $\theta\widetilde<1$ \cite{AlpPin85}. 
\section{GRAVITATIONAL RADIATION FROM PRECESSING PULSARS AND
PULSAR TIMING}
The slowing-down of the pulsar rotation due to gravitational
radiation is limited by the measured value of $\dot P$, so that
$(\dot P)_{\rm GR}<{\dot P}_{\rm obs}$. This
yields observational constraint on the effective tri-axiality of
pulsar,
\begin{eqnarray}
\epsilon_{\rm eff} 
&<& \epsilon_{\rm max}=3\times 10^{-9}
\left({P\over 1~{\rm ms}}\right)^{3\over 2}
\left({\dot P\over 10^{-19}}\right)^{1\over 2}\;,
\nonumber\\
\epsilon_{\rm eff}
 &=& {3\over 8} b {\breve\epsilon} \theta\;.
\label{eq:epsEFFbound}
\end{eqnarray}
We will consider two limiting cases, of a small and large
critical strain, respectively. Numerical estimates for these two
limiting cases, based on the
timing data for two millisecond pulsars and the Crab pulsar, are
 given  in Table {\ref{Table3}}.  
\begin{table}
\caption{Timing data, oblateness, and effective tri-axiality for
pulsars}
\begin{tabular}{cccc}
&&\\
\hline\hline
&&\\
  Pulsar:  & 1937+21 & 1957+20 & Crab(0531+21)\\
&&\\
\hline
&& \\
&& \\
$P$(ms) & 1.56 & 1.61 & 33\\
&&\\
$\dot P$ ($10^{-19}~{\rm s~ s^{-1}}$)& 1.0 & 0.17 & $4.2\times 10^6$\\
&& \\
$\epsilon_{\rm max}$ & $6\times 10^{-9}$ & $2\times 10^{-9}$ &
 $1.2\times 10^{-3}$ \\
&&\\
$\epsilon_\Omega$ & $4\times 10^{-2}$ & $4\times 10^{-2}$ & 
$9\times 10^{-5}$ \\
&&\\
$\epsilon_{\rm eff}$ - case a & $1\times 10^{-10}$ & $1\times 10^{-10}$ & 
$3\times 10^{-13}$\\
 &&\\
$\epsilon_{\rm eff}$ - case b & $4\times 10^{-13}$ & $4\times 10^{-13}$ & 
$3\times 10^{-12}$\\
&& \\
\hline\hline
\end{tabular}
\label{Table3}
\end{table}
In all cases, we assumed rigidity parameter $b=10^{-5}$. 
For the scenario corresponding to `case a', 
we use a conservative assumption of  
$\theta=10^{-3}$. 
In view of the assumed small value of critical strain, we have
in this cas 
$\breve\epsilon\simeq \epsilon_\Omega$. Due to rapid present
rotation of millisecond pulsars, their effective tri-axialities
are, in `case a',  some three orders of magnitude higher than
this of the Crab pulsar. The situation changes in `case b'
(large critical strain). Then, we could contemplate 
the value of $\theta$ for
millisecond pulsars, which could be as large as  1; 
however, due to
the  large critical strain, we should expect
$\breve\epsilon=\epsilon_{\rm init}$, corresponding to $P_{\rm
init}=1~$s, which reduces the effective tri-axiality by some
three orders of magnitude as compared to the `case a'. For the
Crab pulsar we put $\breve\epsilon=\epsilon_{\rm init}$, but in
view of its short age (941 years) we made a conservative choice
of $\theta=10^{-3}$ (no sufficient time to built a large tilt of
the `solid body axis' with respect to rotation axis). Under
these assumptions, we get in  `case b' similar values of
effective tri-axiality for the most rapid millisecond pulsars and
the Crab pulsar. 

The presence of a sufficiently large {\it solid core} would
imply  some difficulties in confrontation with pulsar timing. In
such a (highly speculative) case, we would expect $b\simeq
b_{\rm core}\sim 10^{-2} - 10^{-1}$. Even with a rather small
$\theta=10^{-3}$, the large value of rigidity parameter would lead to
effective tri-axiality largely exceeding the observational upper
bound for millisecond pulsars. Moreover, as the critical strain
for the solid core may be expected to be very large, the
possible `wobble angle' that could be built during star
evolution could, in principle, be large. Confronting these
theoretical speculations with pulsar timing  
leads to conclusion, that if
millisecond pulsars contain a solid core, then their wobble 
angles are extremely small. A more restrictive statement is that
millisecond pulsars are just unlikely to contain a massive solid
core. 
\section{DETECTABILITY}
The radio pulsars with a non-zero tri-axiality  could be
promising sources of continuous gravitational radiation. The
most attractive in this respect are the nearby millisecond
pulsars with $\epsilon_{\rm eff}\sim 10^{-9}$ \cite{New95}.
Consider two selected millisecond  pulsars, PSR 1957+20 and PSR
0437-47. The advantage of the first one is its very short
period, $P=1.6~$ms (the distance being $d = 1530$~pc). On the other
hand, the second millisecond pulsar  is much closer, at $d=140$~pc. 

Millisecond pulsars with non-zero tri-axiality would be
interesting sources of continuous, nearly monochromatic
gravitational radiation, provided some additional conditions are
satisfied. As we could see, the fact that millisecond pulsars
are such  precise clocks ($\dot P<10^{-19}~{\rm s~s^{-1}}$ !),
imposes rather stringent constraints on the maximum value of
$\epsilon_{\rm eff}$. The timing of the millisecond pulsars
tells us, that the possible amplitude of the gravitational waves
should be sufficiently  small to be consistent with observations. 
As a consequence, the detectability of
gravitational waves from the nearby millisecond pulsars by the
forthcoming generation of the LIGO and VIRGO detectors would be
possible only if the periodic signal could be integrated over
the time of a few years \cite{New95}.  This would require a
sufficiently long lifetime of the precession or wobble. 

In the approximation, in which the coupling of the solid and
liquid regions of neutron star are neglected, liquid interior of
neutron star cannot participate in stellar precession. However,
such a coupling exists, and detectability of continuous
gravitational radiation requires that the damping times
connected with the liquid-solid coupling be substantially 
longer than a few
years \cite{AlpPin85}. Finally, the `wobble' is damped 
by the gravitational radiation itself (see Exercise 16.13 in 
\cite{ST83}). Explicit calculation shows that the
characteristic  timescale 
$\tau_{\rm wobble}^{\rm GR}\sim 10^3 
(b\breve\epsilon/10^{-6})^{-2}(P/1~{\rm ms})^{-4}~$years
 (see, e.g.,
\cite{Arau94}): 
the detectability requires, that $\tau_{\rm wobble}^{\rm GR}$ be
substantially longer than a few years. Our discussion of the
damping timescales indicates thus, that the situation in which a
non-zero tri-axiality is due to sufficiently  large, 
genuine `mountains' on the neutron star crust, is the most 
attractive one. This conclusion seems to be in a surprising
harmony with the location of Les Houches School, dominated 
by the magnificent {\it massif du
Mont Blanc},  so admired by all of us during the coffee breaks.
\ack{I am grateful to my wife for a critical reading of the 
manuscript, and for helpful remarks. I am also grateful 
to A.D. Kaminker for preparing Figure 3, and to E. Gourgoulhon 
for reading of the manuscript, and for helpful comments. 
 This work was supported 
in part by the KBN grant 
No. 2 P304 014 07.}
%

%
\end{document}